# Photon radiation calorimetry for anomalous heat generation in NiCu multilayer thin film during hydrogen gas desorption

J. Kasagi[1*], T. Itoh[1,2], Y. Shibasaki[2], T. Takahashi[2], S. Yamauchi[2] and Y. Iwamura[1]

[1] Research Center for Electron Photon Science, Tohoku University, 982-0826 Japan
[2] CLEAN PLANET Inc., 105-0022 Japana

In order to investigate the anomalous heat effect (AHE) in NiCu multilayer thin films, photon radiation calorimetry has been developed. Three types of photon detectors are employed to cover a wide range of wavelengths from 0.3 μm to 5.5 μm, i.e., photon energies from 0.22 to 4.13 eV. In the present work, the usefulness of the calorimetry is demonstrated for excess heat measurements with samples of pure Ni, NiCu composite layers, and Cu mono-layer deposited on a Ni substrate. Direct comparisons of photon radiation spectra with and without $H_2$ easily showed sample-specific differences in excess heat power. The samples of the NiCu composite layer produced larger excess heat. By incorporating the measured radiant power into a heat flow model, the excess heat was deduced to be 4 – 6 W. The energy generated in 80 hours reached 460 $\pm$ 120 kJ: the generated energy per hydrogen was at least 410 $\pm$ 108 keV/H atom. This is definitely not a chemical reaction, but produces energy at the level of nuclear reactions.

## 1. Introduction

After the announcement of "Cold Fusion" by Fleischmann and Pons[1], Ni-based metal + $H_2$ systems have also been studied widely in electrolysis[2-4], as well as by $H_2$ gas[5-7]. However, in experiments using $H_2$ gas that attempted to reproduce high-power production at high-temperature conditions, large excess power as reported was not observed. Regarding the measurements of Levi *et al*[7], Valat *et al* pointed out that they overestimated excess heat by at least one order of magnitude due to errors in emissivity.[8] Budko and Korshunov performed calorimetric measurements in a small experimental setup but no excess heat was observed.[9] As the Google project, Berlinguette *et al* showed that excess power production was not observed for 420 measurements under the conditions of high temperature and high $H_2$ gas pressure with Ni powder including $LiAlH_4$.[10]

On the other hand, in experiments focused on metal nanostructures, Arata and Zhang found that nanosized Pd particles (about 10 nm) with $D_2$ gas generated greater excess heat.[11] Subsequently, Kitamura *et al* confirmed this result by using a flow calorimeter: anomalous large heat production was observed not only with $D_2$ gas but also $H_2$ gas.[12] Furthermore, they expanded the sample from pure Pd nanoparticles to NiPd composite nanoparticles, and also extended the measurement temperature from room temperature to about 600K.[13] In a series of experiments performed as an NEDO project (New Energy and Industrial Development Organization), notable results were obtained as follows: excess power which cannot be explained by chemical reaction was observed up to 600K not only in the PdNi-$D_2$ system but also in the PdNi-$H_2$ system and CuNi-$H_2$ system.[14,15]

We have extended the measurement to a much higher temperature up to 1100 K using nanostructured metal films, instead of the composite amorphous metal powder employed in the NEDO project. The method is rather simple: a nanostructured NiCu multilayer film containing absorbed hydrogen is heated to a high temperature (about 1100K) and evacuated in a vacuum.[16,17] The measurements were performed at a vacuum level of $10^{-4}$ Pa or less, and thus the mean free path of the residual gas in the vacuum is more than 10 m, 100 times the inner diameter of the chamber. Therefore, the heat flow from the sample can be ignored except for thermal radiation and a slight conductive heat flow.

For better evaluation of excess thermal power, we have explored calorimetry by measuring photons emitted from samples.[17] The objective of the present work is to establish photon radiation calorimetry to obtain solid evidence that anomalous large heat production does occur in the nanostructured NiCu multilayer film during desorption of hydrogen gas, and to evaluate the amount of generated heat at high temperatures with high reliability.

## 2. Experiment

### 2.1 Experimental setup

Figure 1(a) briefly shows a top view of the experimental apparatus showing inside of the vacuum chamber. Two thin film samples attached to both sides of a ceramic heater, suspended from the lid of the vacuum chamber, are heated up to about 1100K. Radiant power emitted from the sample surface is measured over a wide range of photon energies with three different detectors by switching two photon analyzers. The data acquired in each detector are combined into one spectrum. This enables not only direct comparison of the radiant power from the sample with and without hydrogen, but also reliable evaluation of the excess heat.

Figure 1(b) shows schematically the sample holder. The sample is a thin foil about 0.2-mm thick. A foil is placed on each side of a ceramic heater (made of alumina, 25 × 25 × 2.5 mm$^3$) which has an R-type thermocouple in the center. The foils are labeled A and B. A 0.3-mm thick insulating plate (Photoveel) is sandwiched between the substrate of the foil and the ceramic heater. The power line of the heater and the reading line of the thermocouple are connected to the outside device via hermetic seals. The foils are covered with two Photoveel plates (40 × 40 mm$^2$ in area with a hole of 20 mmϕ) and fixed to the frame. Photon radiations from the sample foils are measured by three detectors placed outside the windows of the vacuum chamber as shown in Fig. 1(a). For mid-infrared region, two sets of TMHK-CLE1350 (we call Md-IR detector; effective range 0.22 – 0.40 eV) are set at 40 cm from the sample through long vacuum pipes and BaF$_2$ windows. In this experiment, Md-IR is used not as a thermometer, but as a radiation power detector measuring the average radiant power in the 0.22-0.4 eV region.

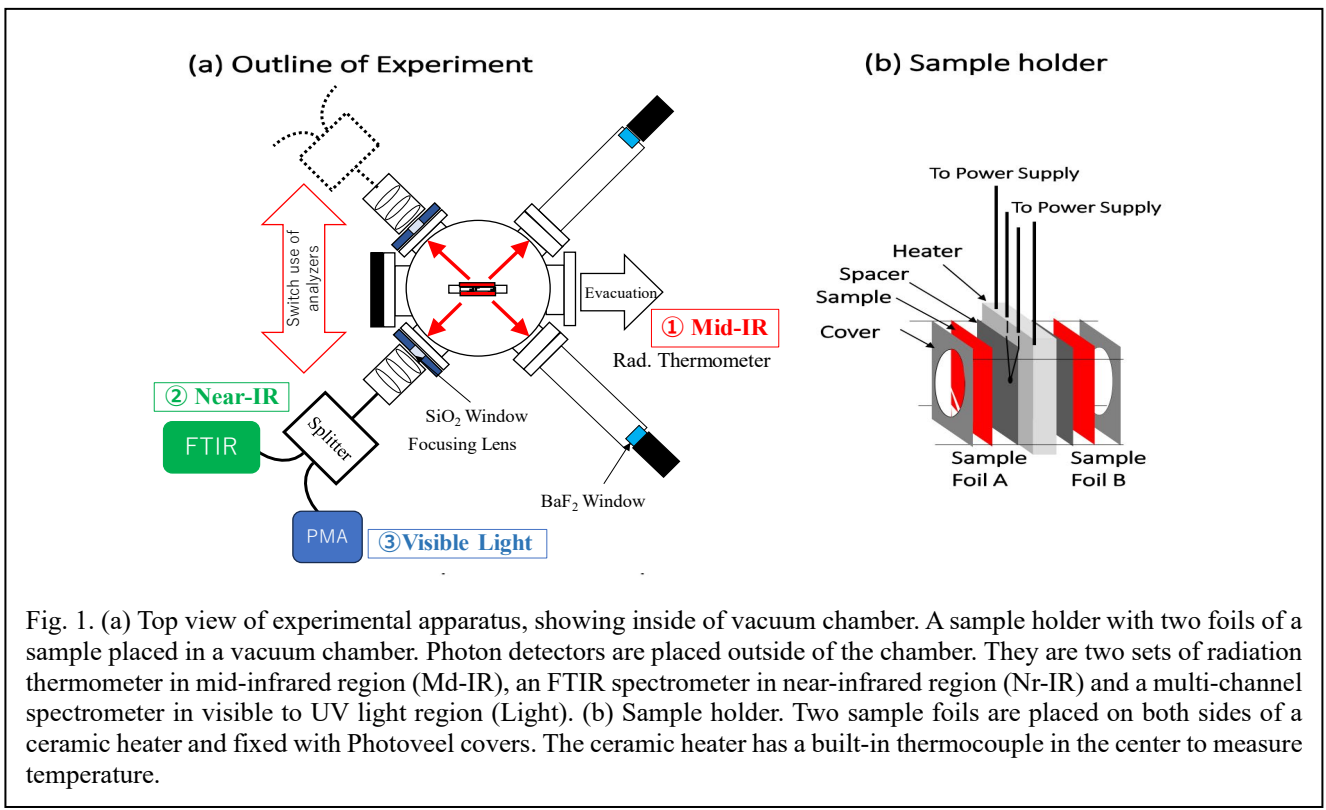

Fig. 1. (a) Top view of experimental apparatus, showing inside of vacuum chamber. A sample holder with two foils of a sample placed in a vacuum chamber. Photon detectors are placed outside of the chamber. They are two sets of radiation thermometer in mid-infrared region (Md-IR), an FTIR spectrometer in near-infrared region (Nr-IR) and a multi-channel spectrometer in visible to UV light region (Light). (b) Sample holder. Two sample foils are placed on both sides of a ceramic heater and fixed with Photoveel covers. The ceramic heater has a built-in thermocouple in the center to measure temperature.

The other two detectors measure the light passing through a quartz window installed in the opposite direction of the long vacuum tube. Behind the quartz window, which is covered with Al foil with a hole of 10 mmϕ on the inner and outer surfaces, the light is focused into an optical fiber cable and then carried by two cables through a splitter. Detectors used are an FTIR spectrometer in the near-infrared region, Hamamatsu C15511-01 (we call Nr-IR detector; effective range 0.5 – 0.9 eV), and a spectroscope in the visible to UV light region, Hamamatsu C10027-01 (we call Light detector; effective range 1.3 – 5 eV).  Outputs of the two detectors are averaged radiation power of multiple measurements as a function of wavelength: measurement time is about 30 seconds for Nr-IR and 50 seconds for Light. It is also noted that measured photons are apertured to accept light only from the samples, not from the entire region including the sample holder.

2.2. Samples

Preparation of samples has been described in detail in [15,16]. Measurements for the following four samples are shown in the present work. (1) Ni pure: 99.9% purity of Ni foil, 0.1-mm thick with area of 25×25 mm$^2$. (2) Ni5Cu1 sample: Layers of Cu (3.8-nm thick) and Ni (20-nm thick) are alternatively deposited by magnetron sputtering to form 6 bilayers on (1) Ni substrate. (3) Ni1Cu3 sample: Layers of Cu (17.8-nm thick) and Ni (6-nm thick) are alternatively deposited by magnetron sputtering to form 6 bilayers. (4) Cu pure: single layer of Cu (140 nm) is deposited on the Ni substrate.

2.3. Experimental Procedure

The measurement was performed for each sample according to the following procedure.

(1) Sample baking: Foils A and B are placed in position in the chamber, which is evacuated down to a pressure less than $3\times10^{-6}$ Pa for a vacuum bake-out by keeping the heater temperature at about 1150K. After about 3 days baking out, proceed to the next step.

(2) Measurement without $H_2$ gas: Keeping inside of the chamber in a vacuum, set the voltage of the heater input ($V_{in}$) to a certain value. Measure the radiant power spectra at about 0.5 h and 3 h after $V_{in}$ is set. Set $V_{in}$ to a different value and repeat the measurement at least 5 different values of $V_{in}$. These measurements are to obtain the reference data corresponding to no excess heat. The results serve for the calibration of null excess energy.

(3) Measurement during desorption of $H_2$ gas: Fill the chamber with $H_2$ gas to 200 – 300 Pa and keep the temperature at about 520K for the sample to absorb the gas. After 12 -15 h, evacuate the vacuum chamber at the same time as setting the value of $V_{in}$. Start the data logger and measure the radiant power spectra at about 0.5, 1.5, 4.0 and 6.0 hours after $V_{in}$ is set. Repeat from the filling of $H_2$ gas for different values of $V_{in}$ at least 4 times. The data will show the temperature dependence of the excess power together with elapsed time dependence for a short period.

(4) Long-term measurement without hydrogen refilling: For the NiCu sample, a long-term measurement is performed in order to see a long-term behavior of the excess power. After hydrogen's absorption into the sample, the measurement with $V_{in}$ = 46 V is started similarly as in (3) but the radiant power spectra are measured every 30 minutes for 5 consecutive days without refilling the $H_2$ gas.

During the above measurements, the data logger records the following information every second to monitor the experiments; heater temperature, output of Md-IR (mid-infrared thermometer), heater voltage and current, vacuum chamber pressure, chamber outer wall temperature.

## 3. Results and discussion

### 3.1. Radiation power spectra

Photon emissions from foils A and B were measured separately as shown in Fig. 1. The two measured intensities were added together to give the average radiant power of the sample. Examples of such radiation spectra are shown in Fig. 2 for samples with different composition ratios. They correspond to the samples, Ni pure, Ni5Cu1, Ni1Cu3 and Cu pure, from left to right, respectively. Black x-marks are the data measured before $H_2$ gas introduced (without $H_2$), and red circles are those measured during $H_2$ gas desorption (with $H_2$). For each pair, the heater input voltage was set to the same value (or input power was almost equal), in order to directly observe the change in radiation power with and without hydrogen.

It is obvious that all samples show enhanced radiation power during $H_2$ desorption, i.e., with $H_2$: Samples with NiCu composite layer show largely enhanced with $H_2$ over without $H_2$, while those with mono-layer of Cu or Ni show slightly.

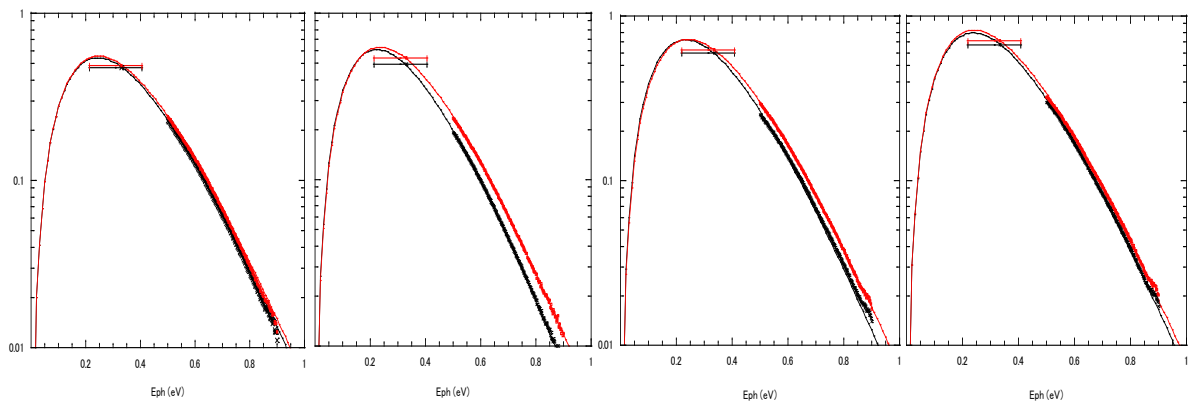

Fig. 2. Examples of photon radiation spectra emitted from samples with different Ni/Cu ratio; Ni pure, Ni5Cu1, Ni1Cu3 and Cu pure, from left to right. Radiant power (W eV$^{-1}$ cm$^{-2}$ sr$^{-1}$) is plotted as a function of photon energy $E_{ph}$ (eV). Red circles are the data measured during $H_2$ desorption, while x-marks are those without $H_2$ gas. For the lowest energy data, the point represents the averaged value within the horizontal bar. A solid line shows the best-fitting calculation with a gray-body radiation approximation.

An important feature of the obtained spectra is that they can be well described by gray-body radiation. Red and black curves shown in Fig. 2 are the best-fitting spectra calculated by gray-body approximation, in which the radiant power is expressed as

$$Y(E_{ph}) = \varepsilon_s (\sigma/\pi) E_{ph}^3/(\exp(E_{ph}/kT_S)-1). \quad (1)$$

Here $E_{ph}$ is the photon energy (eV), $\varepsilon_s$ is the emissivity and $T_S$ is the surface temperature, $\sigma$ is Stephan-Boltzmann constant, and $k$ is Boltzmann constant. Values of $\varepsilon_s$ and $T_S$ are determined so as to give a best fit curve to the data. Table 1 shows the deduced values of $\varepsilon_s$ and $T_S$ corresponding to the spectra shown in Fig. 2. In these examples, $T_S$ increases after H$_2$ introduced, while $\varepsilon_s$ decreases. An increase in temperature is a prerequisite for excess radiant power. The emissivity depends not only on the type of metal but also on the surface condition, and usually decreases mainly due to the reduction of the surface oxide layer by the introduction of H$_2$ gas.

As we will see below, a quantitative evaluation of the thermal power requires the total radiant power flow $Q_S$ from the sample, that is deduced as

$$Q_S = 2\pi A_S \int_0^{\pi/2} \int_0^{\infty} Y(E_{ph}) dE_{ph} \cos\theta \sin\theta \, d\theta = A_S \, \varepsilon_s \, \sigma \, T_s^4, \quad (2)$$

where $A_S$ is the emission area of the sample. Also, the ratio of $\varepsilon_s$ with H$_2$ to without H$_2$ is required to determine the excess power. We define $\beta = \langle\varepsilon_s$ with H$_2\rangle/\langle\varepsilon_s$ without H$_2\rangle$ for each sample, where the numerator is the average value of $\varepsilon_s$ with hydrogen and the denominator is the average value of $\varepsilon_s$ without hydrogen.

Table 1. Sample temperature $T_S$ (K) and emissivity $\varepsilon_s$ deduced from radiation power spectra shown in Fig.1. The last column shows the input power $P_{in}$ (W) during the measurement.

|  | Ni pure<br>with H$_2$/without H$_2$ | Ni5Cu1<br>with H$_2$/without H$_2$ | Ni1Cu3<br>with H$_2$/without H$_2$ | Cu pure<br>with H$_2$/without H$_2$ |
|---|---|---|---|---|
| Ts (K) | 1020 / 1009 | 971 / 926 | 1013 / 984 | 997 / 983 |
| $\varepsilon_s$ | 0.114 / 0.116 | 0.143 / 0.155 | 0.132 / 0.137 | 0.179 / 0.180 |
| Pin (W)* | 34.8 / 35.2 | 34.2 / 35.0 | 25.9 / 26.2 | 28.2 / 28.3 |

*The large difference in heater power between the first two and the latter two is due to changes made to the sample holder to reduce its thermal capacity.

3.2. Evaluation of excess heat

Figure 3 shows heat flow in steady state: a simplified view (left side) and a model diagram used for the analysis (right side). There are two heat sources. One is the heater located between the sample A and B as indicated in Fig. 1(b), and the other is in the sample to generate the excess heat. As shown in the diagram, the heat from the heater ($P_{in}$) transferred to the sample through the spacer and merges with the excess heat ($Q_{ex}$) in the sample. The sum of the heat ($Q_{tot}$) is almost consumed as radiation to the chamber wall from the sample ($Q_S$) and the holder ($Q_H$). Contributions of the conducted heat flow ($Q_C$) through the supporting rod and the reflected radiation from the wall are very small (the latter is ignored in the analysis). Thus, relations between the heat flows and temperatures are expressed by the following equations.

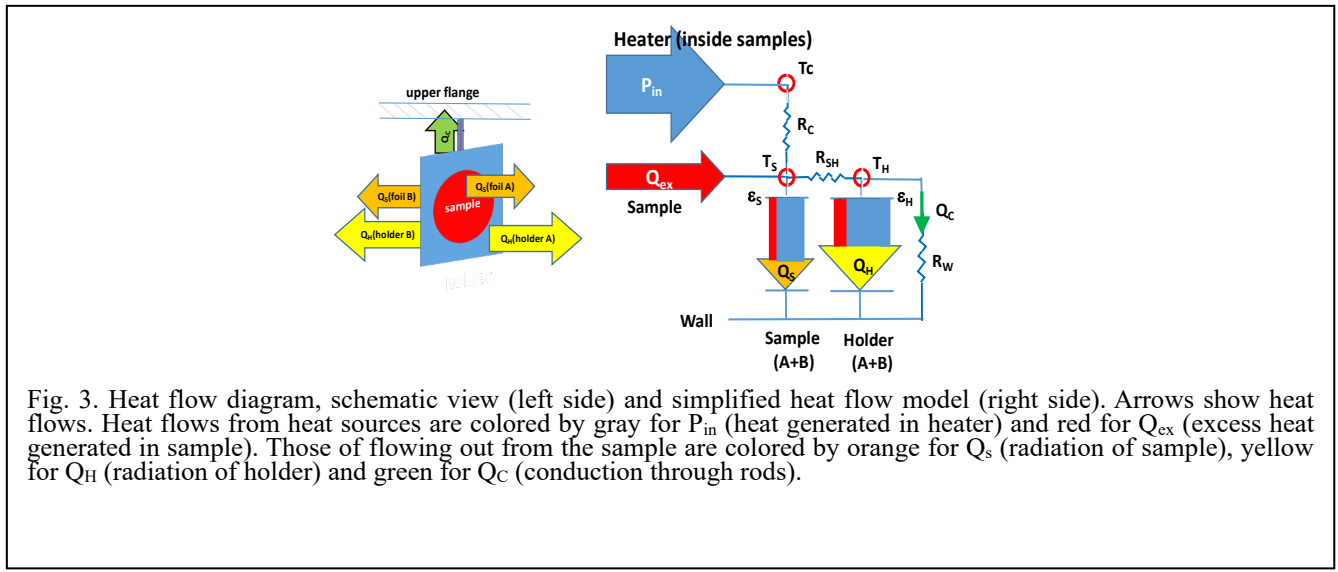

Fig. 3. Heat flow diagram, schematic view (left side) and simplified heat flow model (right side). Arrows show heat flows. Heat flows from heat sources are colored by gray for P$_{in}$ (heat generated in heater) and red for Q$_{ex}$ (excess heat generated in sample). Those of flowing out from the sample are colored by orange for Q$_s$ (radiation of sample), yellow for Q$_H$ (radiation of holder) and green for Q$_C$ (conduction through rods).

$$Q_{tot} = P_{in} + Q_{ex} = Q_S + Q_H + Q_C,$$
$$Q_{S(H)} = A_{S(H)}\, \varepsilon_{S(H)}\, \sigma\, T_{S(H)}^4,$$
$$T_C - T_S = P_{in}\, R_C, \quad (3)$$
$$T_S - T_H = (P_{in} + Q_{ex})\, R_{SH},$$
$$T_H = Q_C\, R_W,$$

where $A_{S(H)}$, $\varepsilon_{S(H)}$ and $T_{S(H)}$ stand for the surface area, emissivity and temperature of the sample (holder), respectively, and $\sigma$ is the Stefan-Boltzmann constant. $R_C$, $R_{SH}$ and $R_W$ are thermal resistances between the heater and the sample, the sample and the holder, and. the holder and the chamber wall, respectively.

Since both $\varepsilon_S$ and $\varepsilon_H$ little depend on temperature in this measurement, $T_S$ and $T_H$ determine the ratio $Q_H/Q_S$ and, hence, it is described as $Q_{tot}$ - $Q_C$ = (1 + $Q_H/Q_S$) $Q_S$. If $R_{SH} = 0$, i.e., $T_H = T_S$, or $T_H = const \times T_S$, then the proportional relation $Q_{tot}$ - $Q_C$ = α $Q_S$ holds strictly (∵ $\alpha = 1 + \frac{\varepsilon_H}{\varepsilon_S}\left(\frac{A_H T_H^4}{A_S T_S^4}\right)$). Even if it is not the case, the numerical simulation shows that linear approximation with $Q_{tot}$ - $Q_C$ = α $Q_S$ + $C$ is effective as long as $T_S/T_H \leq 1.2$ and $Q_S \geq 0.3$ W. Since the estimated value of $T_H$ is about $T_S$ -100 at $T_S \approx 900$K, it is applicable to the present analysis.

However, after introduction of H$_2$ gas, the emissivity of the sample changes to $\varepsilon_S'$ (= $\beta\, \varepsilon_S$), while thermal properties of Photoveel do not change with or without H$_2$; i.e., $\varepsilon_H$ unchanged. In such a case, even when no excess power is generated, $T_S$ increases to $T_S'$ for $\beta < 1.0$. This causes an increase in $T_H$ to $T_H'$ also. But $Q_S$ decreases to $Q_S'$ ($Q_S = A_S\, \varepsilon_S\, \sigma\, T_S^4$ to $Q_S' = A_S\, \varepsilon_S'\, \sigma\, T_S'^4$) by the amount that $Q_H$ increases to $Q_H'$ ($Q_H = A_H\, \varepsilon_H\, \sigma\, T_H^4$ to $Q_H' = A_H\, \varepsilon_H\, \sigma\, T_H'^4$), since $Q_{tot}$ stays constant. In other words, after hydrogen is introduced, the term $Q_H/Q_S$ becomes $Q_H'/Q_S'$ (> $Q_H/Q_S$) and the value of α determined before H$_2$ introduction should be corrected corresponding to this change. Using the approximation, $T_H' = const \times T_S'$, we replace α with $\alpha' = (\alpha + \beta - 1)/\beta$. Thus, the value of $Q_{tot}$ must be evaluated by the following calibration equation.

$$Q_{tot} = P_{in} + Q_{ex} = \alpha'\, Q_S + Q_C + C, \quad (4)$$
where $\alpha' = (\alpha + \beta - 1)/\beta$ with $\beta = \varepsilon_S'/\varepsilon_S$.

Although the conductive heat flow $Q_C$ is very small as around 0.13 W, we evaluate it as $Q_C = -3.71 \times 10^{-3} - 3.603 \times 10^{-5} T_C + 1.257 \times 10^{-7} T_C^2$, which was experimentally determined before the measurement.

Fig. 4 shows examples of calibration curves, $P_{in}$ - $Q_C$ vs $Q_S$ for Ni5Cu1 and Ni pure samples; they are measured for various $P_{in}$ before the introduction of H$_2$ gas. Since no excess heat is expected without H$_2$ gas, we deduce values of α (corresponding to $\beta = 1$) and $C$ by linearly fitting to the data. The solid blue circles are for the Ni5Cu1 sample and the open circles for the Ni pure sample. It is shown that the two data sets are very well fitted with respective straight lines. It should be noticed that the calibration curve is slightly different for each sample. This may be due to the individuality of the samples and subtle differences in the setting conditions.

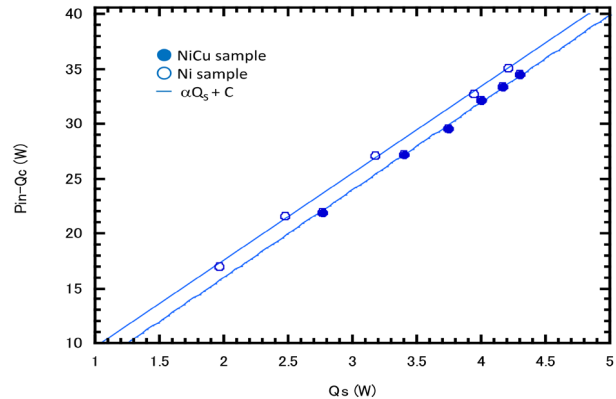

Fig. 4. Calibration curves for Ni pure sample (open circles) and for Ni5Cu1 sample (closed circles). Data before H$_2$ gas introduction are plotted; Input power of the heater against deduced values of Qs.

The evaluation of $Q_{ex}$ by the present calorimetry using measured $Q_S$ and Eq. (4) may be more reliable than the conventional method that only uses heater temperature $T_C$. The following two points must be noted. (1) The effect of change in $\varepsilon_s$ of the sample due to H$_2$ gas introduction should be taken in account: When $\varepsilon_s$ decreases by 10% due to hydrogen introduction, $T_S$ increases by about 2%, even without excess power generation. This results in an increase in $T_C$ by similar amount. (2) The heat flow from the heater can be directly used as calibration for the present calorimetry: As shown in Fig. 3, the heat flow from the heater ($P_{in}$) determines the temperature difference $\Delta T$ between $T_C$ and $T_S$ and the temperature of $T_S$, but the heat flow ($Q_{ex}$) generated in the sample does not contribute to $\Delta T$. Thus, it is not correct to use the relationship between $T_C$ and $P_{in}$ as a calibration curve, since this requires a situation that $Q_{ex}$ originates at the heater position.

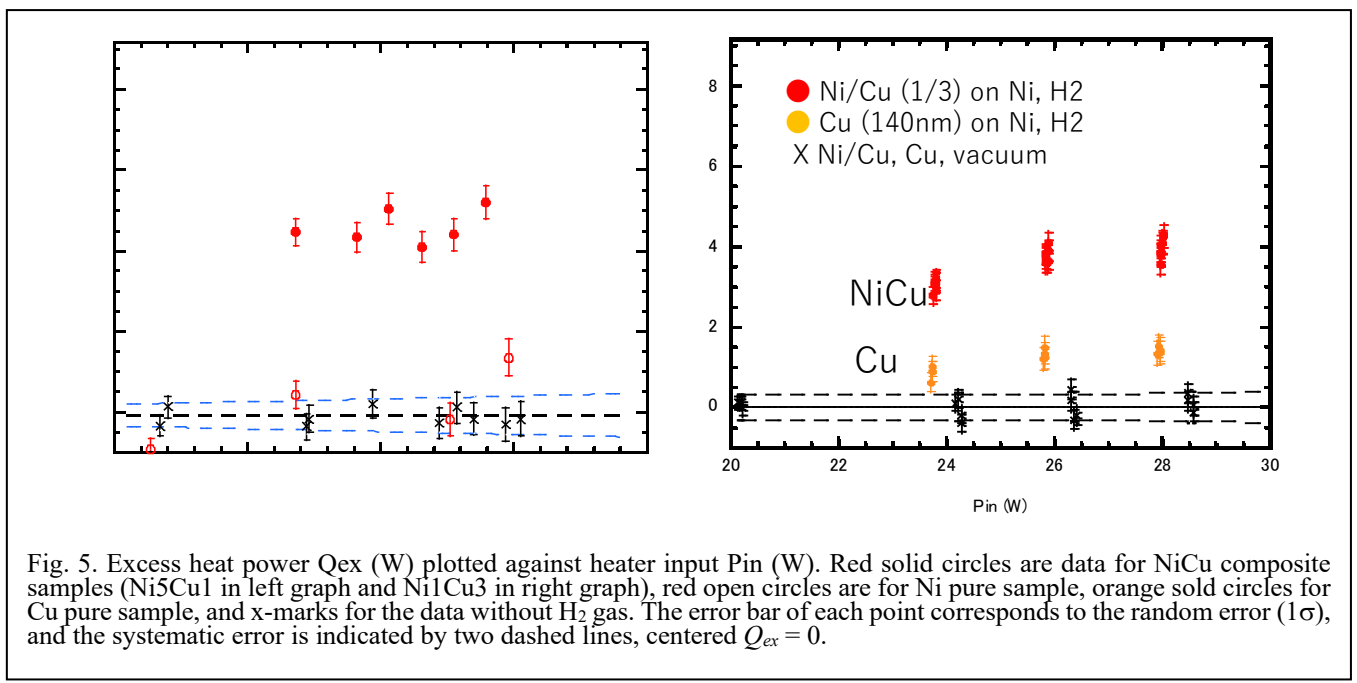

Fig. 5. Excess heat power Qex (W) plotted against heater input Pin (W). Red solid circles are data for NiCu composite samples (Ni5Cu1 in left graph and Ni1Cu3 in right graph), red open circles are for Ni pure sample, orange sold circles for Cu pure sample, and x-marks for the data without H$_2$ gas. The error bar of each point corresponds to the random error (1σ), and the systematic error is indicated by two dashed lines, centered $Q_{ex} = 0$.

## 3.3. Quantitative comparison of excess power

Applying Eq. (4), we have obtained the excess heat power $Q_{ex}$, shown in Fig. 5 as a function of the heater input power $P_{in}$; data of NiCu composite samples (Ni5Cu1 in left and Ni1Cu3 in right) are plotted by red solid circles, those Ni pure sample are plotted with red open circles in left, and those Cu pure are plotted with orange closed circles. Data without $H_2$ used for the calibration curve are plotted with black x-marks. The error bar of each point corresponds to the random error (1σ), mostly originated from the error of $Q_S$, roughly 0.18% of its value. Two dashed lines, centered $Q_{ex} = 0$, show the systematic error (1σ) for $Q_{ex}$, being due to the uncertainty of the parameters of the calibration ($\alpha$ and $C$) and the ratio of the emissivity ($\beta$). It is shown clearly that the NiCu composite sample with $H_2$ produces large excess heat power; the amount is about 4 - 5 W and up to 17% of $P_{in}$. In contrast, the result of the Ni pure sample with $H_2$ is almost indistinguishable from those without $H_2$. The Cu pure sample slightly but surely produces $Q_{ex}$. However, it is not certain that the excess energy source is in the Cu layer, since a possibility of alloying with Ni substrate cannot be ruled out. Using the thermal diffusion coefficient of Cu in Ni [18], all of the Cu on the surface is dissolved into the Ni substrate.

It is clear that the Ni/Cu binary thin film plays an important role in generating the large heat. However, the role of the initial layer structure may not be significant, since the layer structure may be lost thoroughly during the 3-day baking at about 1150K before the measurement start. Measurements of the depth distribution of elements in samples after the excess heat measurement have been reported in [16] for composite thin films Ni/Cu and Ni/Cu/$Y_2O_3$. In the Ni/Cu sample, no traces of a layered structure were observed, while in the Ni/Cu/$Y_2O_3$ sample, decreasing wavy distribution of Y concentration was observed near the surface. The relationship between the magnitude of the generated heat power and the Ni/Cu ratio as well as the Ni/Cu structure remains to be addressed in the future.

## 3.4. Produced energy

Fig. 6 shows an example of long-term continuous measurement of $Q_{ex}$. The total energy generated in 80 hours is obtained by summing up the experimental values to be 460 ± 120 kJ. Since this energy is excessively generated after $H_2$ is occluded in the sample, the energy source is considered to be a rection involving hydrogen in the sample. For the measurement of Fig. 6, hydrogen molecules of $6 \times 10^{-6}$ moles were absorbed in the sample. Then, the amount of energy per H atom is 410 ± 108 keV/H atom, under the assumption that all H atoms are involved in the reaction. This is huge amounts of energy that can never be produced from energy exchanges in electronic levels. The actual amount of the produced energy/H should be much larger, when considering the fact that most of hydrogens initially absorbed discharged into a vacuum by thermal diffusion very soon, without involving in the reaction.

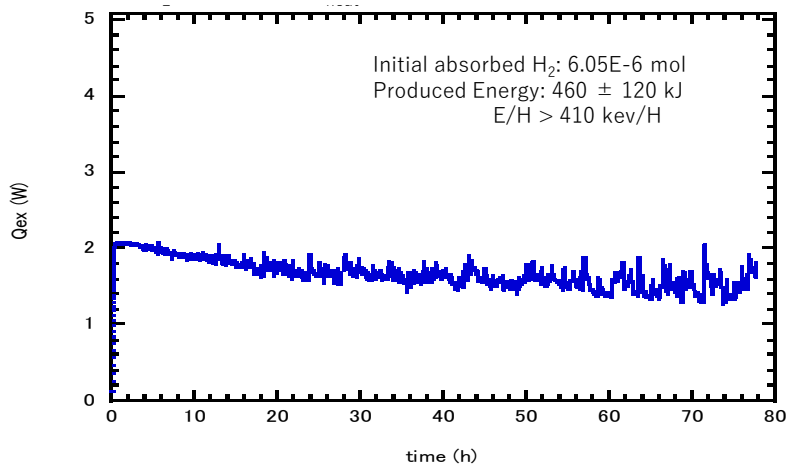

Fig. 6. A result of 80-hour continuous measurement with two Mid-IR detectors for the Ni1Cu3 sample. Deduced excess heat power $Q_{ex}$ (W) is plotted as a function of elapsed time (hour).

## 4. Conclusion

For studies of excess heat generation in NiCu multilayer thin film with $H_2$ gas, the photon radiation calorimetry has been established. Three types of photon detectors employed in the measurement give a radiation intensity spectrum covering a wide range of photon energies from 0.2 to 1.8 eV. This serves to compare the excess power produced in the sample, visibly although qualitatively. The spectrum is well approximated by a gray-body radiation; the emissivity and temperature of the sample can be deduced. This is important, because a change in thermal properties may affect the evaluation of the excess power, especially at high temperature like present case (up to 1100K). It is emphasized that the radiant calorimetry has advantageous over thermometry.

By incorporating the measured radiation power into the heat flow model, one can evaluate the excess heat power quantitatively. It is emphasized that the calibration can be made correctly and the effect of change in emissivity can be taken into account. It was found that the sample having NiCu composite layer always produced larger excess heat than Ni (Cu) pure sample; the excess heat was deduced to be 4 – 6 W. The energy generated in 80 hours reached to 460 ± 120 kJ: the generated energy per hydrogen was at least 410 ± 108 keV/H atom. This is definitely not a chemical reaction, but producing energy at the level of nuclear reactions.

## Acknowledgements
This work is supported by Clean Planet Inc., and The Thermal & Electric Energy Technology Foundation.